\documentclass[nofootinbib]{nature}

\usepackage{amsmath}    
\usepackage{color}
\usepackage[10pt]{moresize}

\usepackage{graphicx}
\usepackage{caption}
\DeclareCaptionLabelFormat{adja-page}{\hrulefill\\#1 #2 \emph{(previous page)}}
\usepackage{subfig}

\usepackage{amsfonts}
\usepackage{amssymb}
\usepackage{amscd}
\usepackage{amsmath}
\usepackage{enumerate}
\usepackage{epsfig}
\usepackage{epstopdf}
\usepackage{dcolumn}
\usepackage{bm}
\usepackage{amsthm}
\usepackage{amsfonts}
\usepackage{color} 
\usepackage[usenames,dvipsnames]{xcolor}
\usepackage{amsmath}
\usepackage{enumerate}
\usepackage{bbm}
\usepackage[colorlinks=true,citecolor=blue,urlcolor=black]{hyperref}

\usepackage{mathrsfs}

\begin{document}

\title{Single photon imaging and sensing of obscured objects around the corner}

\author{Shenyu Zhu$^{1,2}$, Yong Meng Sua$^{1,2}$, Patrick Rehain$^{1,2}$, Yu-Ping Huang$^{1,2}$}

\maketitle

\begin{affiliations}
 \item Department of Physics, Stevens Institute of Technology, 1 Castle Point Terrace, Hoboken, New Jersey 07030, USA
 \item Center for Quantum Science and Engineering, Stevens Institute of Technology, 1 Castle Point Terrace, Hoboken, New Jersey  07030, USA
\end{affiliations}

\maketitle

\begin{abstract}
Non-line-of-sight (NLOS) optical imaging and sensing of objects imply new capabilities valuable to autonomous technology, machine vision, and other applications. Existing NLOS imaging methods rely heavily on the prowess of computational algorithms to reconstruct the images from weak triply scattered signals. Here, we introduce a new approach to NLOS imaging and sensing using the picosecond gated single photon detection generated by quantum frequency conversion. With exceptional signal isolation, this approach can reliably sense obscured objects around the corner and substantially simplify the data processing needed for position retrieval and surface profiling. For each pixel, only $4 \times 10^{-3}$ photons are needed to be detected per pulse to position and profile occluded objects with high resolution. Furthermore, the vibration frequencies of different objects can be resolved by analyzing the photon number fluctuation received within in a ten-picosecond window, allowing NLOS acoustic sensing. Our results highlight the prospect of photon efficient NLOS imaging and sensing for real-world applications.

\end{abstract}

\section*{Introduction}

The capacity of optical detection and imaging technology is ever expanding to keep pace with emerging autonomous technology and evolving sensing needs. In particular, the desire to see around corners has attracted much research interest from various fields, with the prospect of unlocking new imaging modalities over a breadth of applications, such as non-line-of-sight (NLOS) imaging and NLOS tracking for machine vision and sensing, autonomous driving, and biomedical imaging \cite{maeda2019recent}. The ability to sense, track, and image occluded objects with sufficient resolution and accuracy is valuable for autonomous technology and machine vision when direct line-of-sight is prohibited or split second decision-making is needed for preemptive safety measures \cite{Isogawa_2020_CVPR, Zubairu2018}. Practical NLOS imaging and sensing is an interdisciplinary problem at the intersection of physics, optics, and signal processing. It requires a sophisticated optical measurement system for capturing information-carrying photons, combined with an appropriate light transport model for efficient and reliable reconstruction of hidden scenes with reasonable computational overhead. 

Over the last decade, we witnessed considerable progress in NLOS imaging and sensing based on advanced measurement systems such as streak cameras \cite{Velten2012}, single-photon sensitive avalanche diodes (SPADs) \cite{Gariepy2015, liu2020phasor, chan2017non, liu2019non, brooks2019single}, and interferometric detection \cite{lei2019direct, batarseh2018passive, klein2016tracking, metzler2020deep}. Leveraging their high sensitivity, a diverse toolbox of NLOS reconstruction algorithms have been developed based on various light transport models for recovering hidden scenes \cite{Velten2012, O'Toole2018, liu2020phasor, iseringhausen2020non}. Notwithstanding, those methods, maybe with few exceptions, require a priori information on the hidden scene and are hence overly restricted to imaging the hidden and obscured objects of known geometry. Arguably, all those NLOS reconstruction approaches rely on faithful measurements of the optical signals that carry the information of the scene.   

In a typical NLOS scenario, the probe laser beam first bounces off a single point on a diffusive wall, with some photons redirected towards the hidden scene. A small portion of those photons are back-scattered by the scene and redirected again by the wall to reach a detector. The detector can consist of a separate receiver, which captures photons from a different point on the wall. It can also use a transceiver to capture those from the same point, for coaxial NLOS measurement. In either case, as the intensity of light scattered from a diffusive surface is bounded by the inverse-square (distance) law, in such a NLOS scenario, those triply-bounced information-carrying photons decay several orders of magnitude faster amid much brighter photons returning directly from the wall. Many existing optical NLOS imaging and NLOS tracking systems, achieved by a single pixel \cite{PhysRevApplied.12.011002, chan2017non, liu2019non} or 2D \cite{Velten2012, Gariepy2015, liu2020phasor} single-photon detector, capture the back-scattered photons using a separate receiver to avoid receiving the photons directly returning from the wall, which may saturate the single photon detector and suffer the pile-up effect \cite{heide2018sub}. However, NLOS with a separate receiver has to aggressively illuminate and image pairs of distinct points on the wall for the time resolved single photon detection \cite{Wue2024468118}. On the other hand, coaxial NLOS systems which uses a monostatic single transceiver setup, benefit from a straightforward geometrical relation of the time-of-flight measurement and the hidden scene. It can utilize simpler algorithms with much less computational complexity, such as light-cone transformation, to reconstruct the scene \cite{O'Toole2018, lindell2019wave}. The drawback, however, is the strong pile-up effect. To avoid this issue, in the previous confocal NLOS setup, the targeted object was placed far away from the wall, and the receiving field-of-view of the SPAD was carefully aligned to be slightly off the illumination point of the outgoing probe beam on the wall \cite{O'Toole2018}. Also, there was no obscurant between the targeted object and the wall, because any obstacle in front of the object will further attenuate the information-carrying photons while also increasing background photons that are hard to be rejected based on time-of-flight, rendering complications in reconstructing the hidden scene \cite{Thrampoulidis2019, kadambi2016occluded, FelixHeide2019NLOSOcclude, xu2018revealing, rapp2020nloserti}. The above restrictions pose a significant challenge in practical NLOS imaging and sensing, preventing them from deployment with complex scenes and possible presences of obscurants. 

Here, we aim at overcoming this challenge by distinguishing the information-carrying photons from overwhelming background photons in a coaxial NLOS setting and introduce a new optical detection modality for NLOS imaging and sensing. We demonstrate a single pixel NLOS imaging and sensing system based on time-correlated single-photon counting through nonlinear optical gating \cite{Shahverdi:18}. It achieves an absolute 10 ps temporal resolution for object imaging, positioning, and surface normal retrieval, as well as vibration sensing of highly obscured objects around the corner. Our method employs highly-efficient and low-noise quantum frequency conversion of single photons in a nonlinear waveguide, where a time correlated 6-ps pump pulse performs effectively as a narrow nonlinear optical gating to up-convert 6-ps signal photon via sum-frequency generation (see supplementary 1). Crucially, the pump pulses create a high extinction picosecond photon detection window much narrower than timing jitter of the detector and its associated electronics, thus realizing a mechanism to isolate and distinguish information-carrying NLOS photons from the background photons that are usually several orders of magnitude stronger \cite{Shahverdi:18, Shahverdi2017}. This picosecond photon detection window will minimize the photon count distortion from pile-up and detector saturation, which are otherwise plaguing applications based on conventional single photon detection \cite{Heide2018, Maruca:21}. The system thus provides the ability of picosecond-precision NLOS objects recovery even in highly obscured scenarios. The capability of exclusively capturing the photons from the target open the path towards NLOS vibration sensing, which will underpin the prospect of new hybrid imaging modalities, such as acousto-optics imaging or photoacoustic remote sensing, for NLOS application \cite{lindell2019acoustic}.

\begin{figure}\center
\includegraphics[width=15.9 cm]{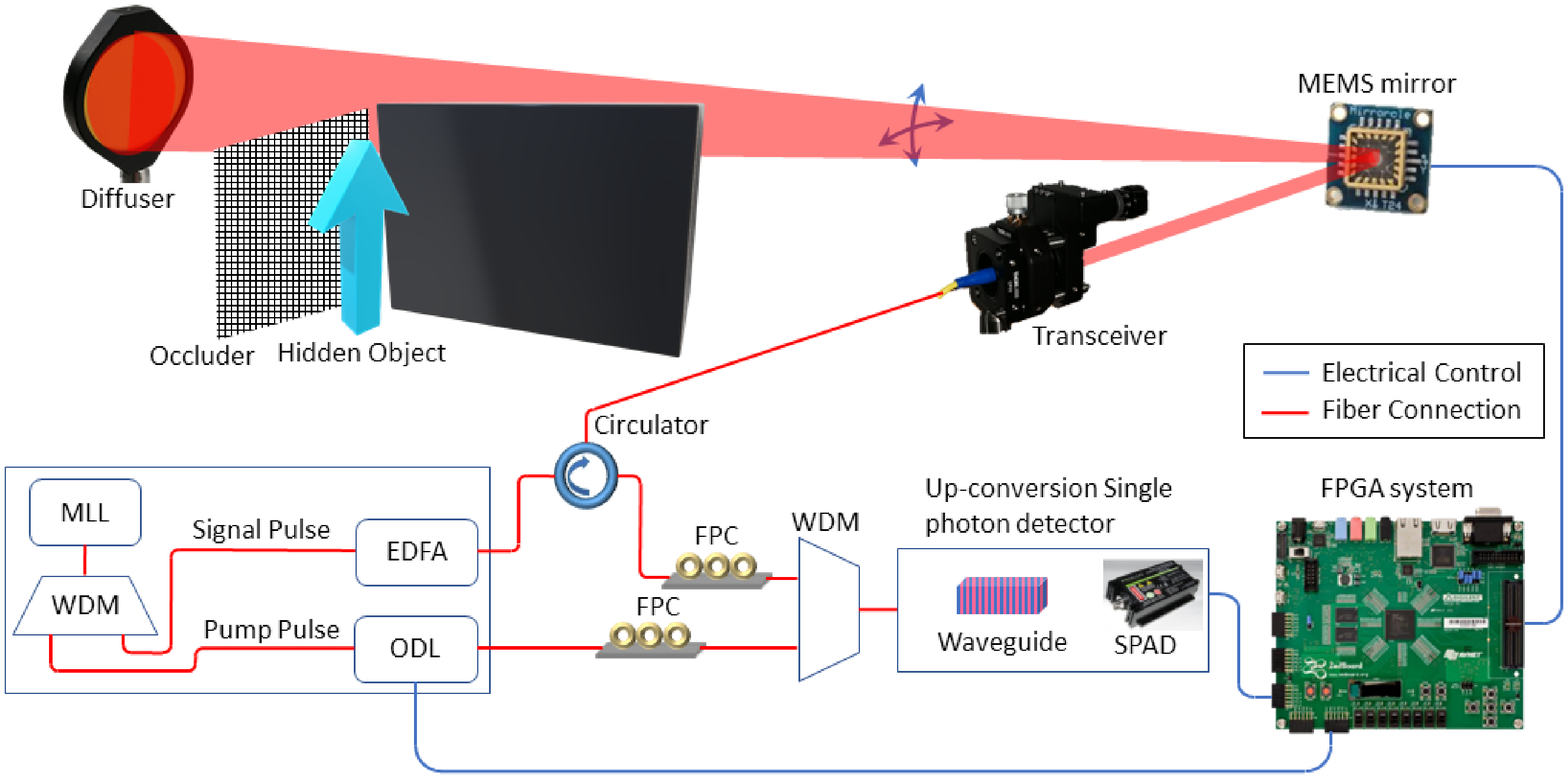}
\caption{\textbf{A sketch for the NLOS system.} MLL: mode lock laser; WDM: wavelength division multiplexer, which is used as optical filter just after the MLL, and as combiner before going into the nonlinear optical gating detector; ODL: optical delay line; FPC: fiber polarization controller. The nonlinear gated single photon detector contains a quasi phase-matched nonlinear waveguide module and a silicon SPAD. The system transmits probe laser and receives signal photons using the transceiver, then the hidden object obscured by the aluminum mesh is imaged and sensed.
}
\label{figure_sketch}
\end{figure}

The proof-of-principle experiments demonstrating NLOS imaging, positioning and sensing of highly obscured objects are shown in Fig.~\ref{figure_sketch}. The setup consists of a mode-locked laser (MLL), a micro-electromechanical-system (MEMS) scanning mirror, a single-mode fiber (SMF) coaxial optical transceiver, a programmable optical delay line (ODL) and a silicon SPAD. The scene around the corner is realized by using a 2-inch diameter metallic diffuser as the wall, an aluminum mesh (1 mm diameter wire grid with 2 x 2 mm openings) as the obscurant before the hidden object. A probe pulse train derived from the MLL is collimated and sent out via the transceiver, and steered by the MEMS mirror to perform 2D raster scan over different points on the diffuser. After triply-bounced, few of the information-carrying NLOS photons are scattered back in the retro direction, and coupled into the coaxial transceiver. On the other hand, the pump pulse train, derived from the same MLL and synchronous with the probe, is sent through the ODL for temporal scan along the depth dimension to facilitate time-resolved photon counting with high resolution. It is combined with the received photons in a dense wavelength-division multiplexer (DWDM) into a quasi phase-matched nonlinear waveguide for frequency up-conversion. Only when the received photons are temporally aligned with the pump can the up-conversion process achieve high efficiency. Subsequently, the up-converted photons are detected by a silicon SPAD. The entire system thus realizes nonlinear gated single photon detection (NGSPD) \cite{rehain2020noise}, which distinguishes the information-carrying NLOS photons while rejecting the photons scattered back directly from the diffuser or obscurant even in this coaxial transceiver setup.


\section*{NLOS imaging of highly obscured target}

To probe and image the targeted scene, the MEMS scanning mirror steers the probe laser beam for raster scanning 32$\times$32 points on the diffuser, while recording the photon count as a function of temporal delay of the pump at each scanning point. This results in a temporally resolved 3-dimensional photon count array whose axes are $x, y$ (scanning coordinates on the diffuser) and $t$ (relative temporal delay of the pump).  

\begin{figure}[!t]\center
\includegraphics[width=15.9 cm]{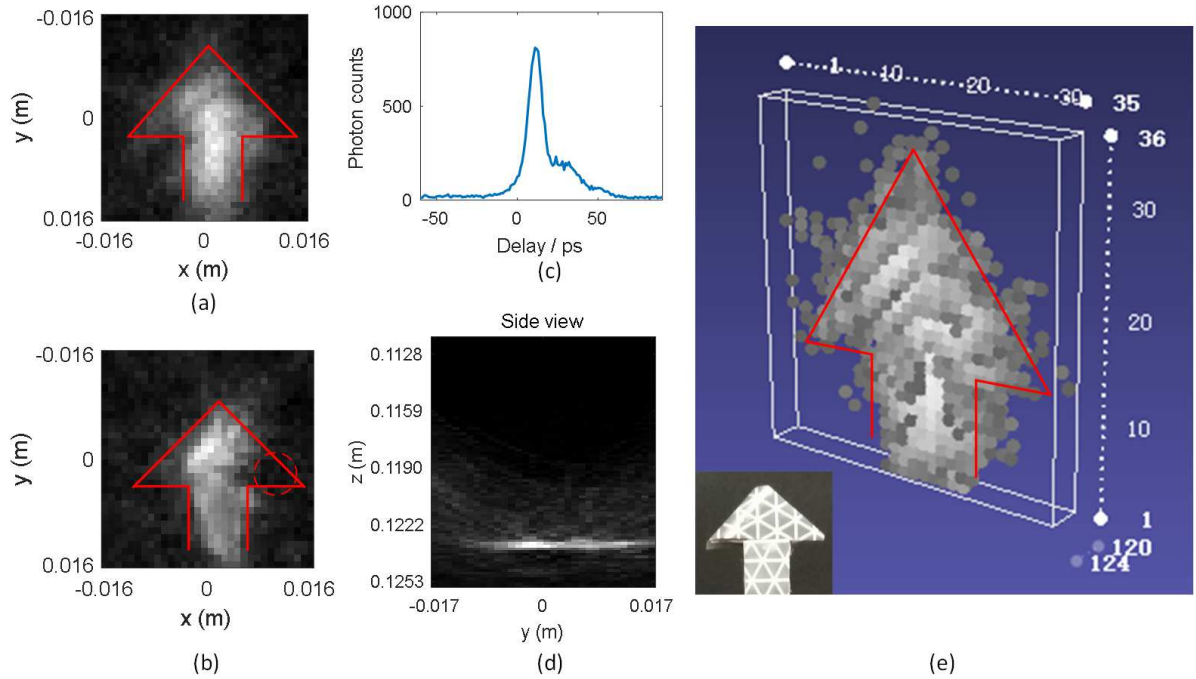}
\caption{\textbf{Imaging results of NLOS imaging} Imaging result of a retroreflective arrow. (a) and (b) are the front views of the reconstructed NLOS imaging result for no obscurant and with obscurant, with red lines indicating the profile of the arrow. Most of the arrow shape is reconstructed with the existence of obscurant, except for the right arrow tip labeled with dashed red circle. The photo of the arrow is at the bottom-left corner of (e). (c) is an example of the time-histogram for the obscured object, and (d) is the side view of the NLOS reconstructed result with the obscurant, which shows ps-level depth resolution of the surface. Note that few counts from the obscurant was received by the single mode fiber transceiver, which explains the very weak response from the obscurant itself lies in the front of the arrow. (e) renders a 3D point-cloud of the reconstructed result (no obscurant) using MeshLab\cite{CignoniConf2008}, which is labeled in millimeter.  
}
\label{figure_result_image}
\end{figure}

The 3-dimensional photon count array is then processed to reconstruct the NLOS scene. Prior to image reconstruction, we pre-process the raw data by first compensating the relative time-of-flight difference caused by the tilt angle of the diffuser - since the time-of-flight from the transceiver to different scanning points on the diffuser varies with optical path (see supplementary Note 3). Then the time-resolved photon counting histogram at each scanning point is filtered individually using a one-dimensional convex target function
\begin{equation}
f = \mathop{argmin}\limits_{\textbf{x}}  (\|\textbf{y} - \textbf{Ax} - \textbf{e}\|_{2} + \lambda \| \textbf{x} \|_{1})
\end{equation}
with CVX toolbox \cite{cvx, gb08} for Matlab. In the target function above, \textbf{y} is the time-resolved histogram measurement (Fig.\ref{figure_result_image}(c) as an example) on one scanning point, \textbf{e} is the average background noise level, \textbf{x} is the filtered time-resolved histogram(target), $A$ is the impulse response matrix of single point object, where the impulse response of the system is measured to be 10 ps FWHM (see supplementary Fig.2). This optimization procedure has a similar form of compressive sensing recovery, which removes the background noise due to the intrinsic dark count of the NGSPD and the ambient light. $\lambda \| \textbf{x} \|_{1}$ is added as a $l_{1}$ regularizer to prevent over-fitting of the processed data, and $\lambda$ is set at a low value(0.1) to preserve the signal response thus not overly sparsifying the target. Subsequently, the targeted scene can be recovered from the processed data by using the 3-dimensional reconstruction algorithm based on light-cone transformation \cite{O'Toole2018}. 

A typical retroreflective \cite{O'Toole2018} arrowhead is used as the imaging target, shown in the inset of Fig.\ref{figure_result_image} (e). The arrowhead is positioned at 12 cm in front of the diffuser, where its line-of-sight to the transceiver is blocked. We first perform the NLOS imaging as is, and afterward insert the obscurant at about 1 cm right in front of the target. The obscurant reduces considerable amount of the information-carrying photons from the target while inducing substantial back-scattered photon ahead of them, thus likely to conceal the target from non-gated single photon detection \cite{Heide2018, O'Toole2018}. Utilizing NGSPD to negate the drawbacks due to the obscurant, we are able to reconstruct the image of the NLOS arrowhead behind the obscuring aluminum mesh in high accordance to the arrowhead as shown in Fig.\ref{figure_result_image}(b), which maintains most of the image features compared to the front view and 3D point-cloud of the ground truth under the same illuminating probe power, shown in Fig.\ref{figure_result_image} (a) and (e). This is due to few-picosecond gate and timing resolution of nonlinear gating, as the NGSPD NLOS imaging system can distinguish the back-scattered photons from the diffuser and the obscurant despite using a coaxial transceiver. The NGSPD system can distinguish the diffuser and the object even if they are separated by only 1 cm (See Supplementary Fig.S3). The other reason for this obscurant-rejection is due to the SMF coupled coaxial optical transceiver's point spread function (PSF) on the diffuser is equivalent to a spatial filter which prevents back-scattered photons from the obscurant from overwhelming the detector. Even though the probe pulse is diffusely illuminating the target and obscurant, only back-scattered photons falling into the transceiver's PSF on the diffuser will be detected and captured in the time-resolved histogram. The existence of the obscurant increases the background count slightly while reduces considerable amount of detected photon thus deteriorated the reconstruction quality compared to ground truth, as the tip of arrow (labeled with red dashed line circle) was not manifested in Fig.\ref{figure_result_image}(b). 

Considering that the temporal resolution of the NGSPD ${\Delta}t \approx 10 ps$, the spatial resolution of this coaxial NLOS imaging system based on NGSPD can be estimated as ${\Delta}w=\frac{c\sqrt{w^2+z^2}}{2w}{\Delta}t \approx 1.1cm $ \cite{O'Toole2018}, where $z$ is the distance from the diffuser to the object, $w$ is half of the spatial scanning range on the diffuser and ${\Delta}t$ is the temporal resolution. As the size of the arrow is in centimeter scale, it is remarkably well resolved in the reconstructed images except at the sharp tips of the arrow with feature size well below 1 cm. The total acquisition time for one image is about 15 minutes at a rate of 10 ms dwell time per delay point. For reconstructing Fig.\ref{figure_result_image}(b), only about $4 \times 10^{-3}$ detected information-carrying NLOS photon counts per pulse per pixel is required at the peak of the time-resolved histograms, thanks to the very low noise of NGSPD. The photon efficient nature of NGSPD points to its expectable potential for direct single photon imaging at long distance or NLOS single photon imaging where in these scenario the returning photons are very rare \cite{Li:21,Wue2024468118}. 

\section*{NLOS position and orientation retrieval of obscured targets}

Identifying the position and surface normal of the obscured NLOS target requires the capability of isolating or being able to identify the information-carrying photons from the target rather than obscurant \cite{Gariepy2015, brooks2019single}. This can be achieved via NGSPD, by acquiring pristine and picosecond-resolved photons arrival time-resolved histogram. 

\begin{figure}[!t]\center
\includegraphics[width=12.0 cm]{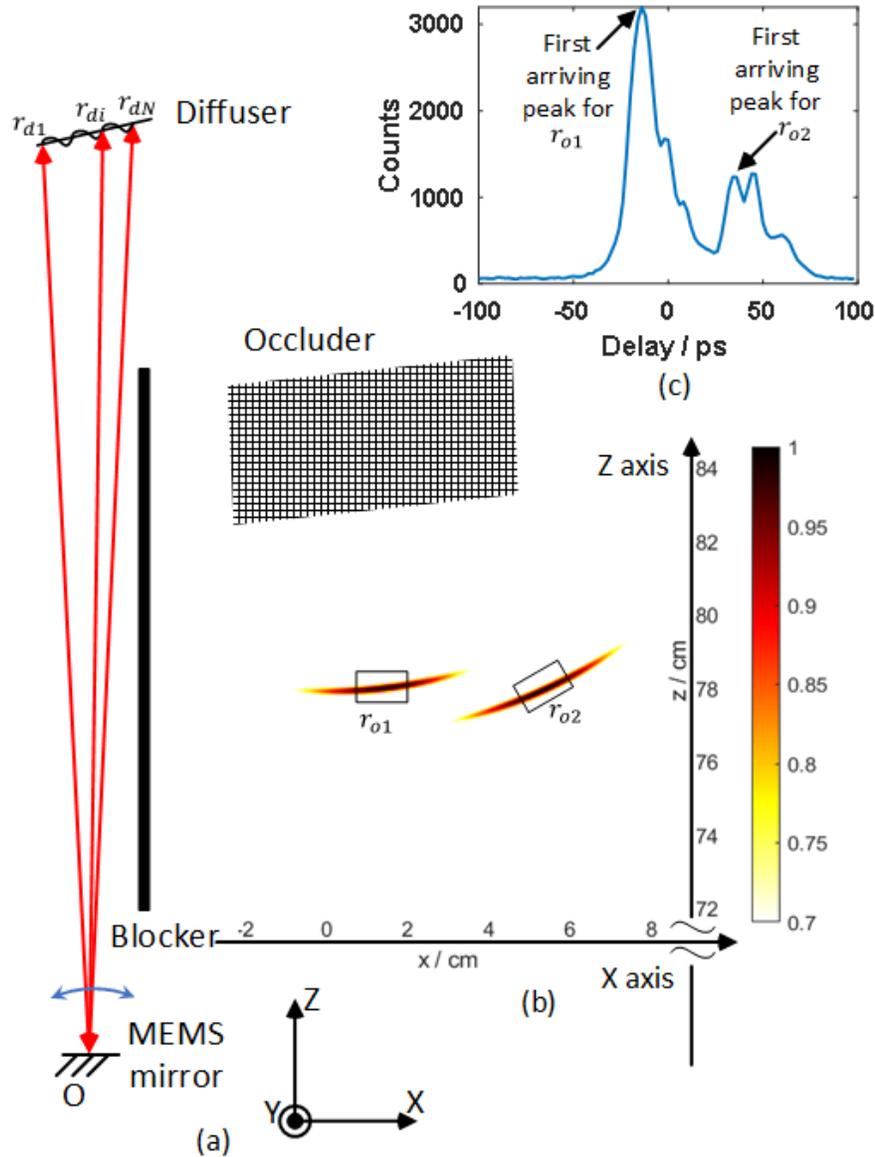}
\caption{\textbf{NLOS target's position retrieval} (a) The setup sketch top view of the NLOS positioning through obscurant, where the distance from the MEMS mirror to the diffuser is 90 cm. (b) is the least-sum-square fitting result of the x-z coordinate of the two bars, where the color scale indicates the probability where the bars stand. The x-z plane is originated at the MEMS mirror. For better seeing the result, the reconstruction is not in the same scale of the setup sketch. (c) The time resolved measurement on one pixel shows different time-of-flight result from the two bars.}
\label{figure_result_position}
\end{figure}

In this experiment, we place two 4-cm distanced retroreflective bars that are both about 12 cm in front of the diffuser, and having the obscurant in between, shown in Fig.\ref{figure_result_position}(a). We scan the probe on the diffuser along a single horizontal row of points and record the photon arrival time-resolved histogram. The NGSPD temporally resolves the back-scattered NLOS photons from different objects with small time-of-flight difference, which enables retrieving each bar's position. An example of the time-resolved histogram at one scanning point is shown in Fig.\ref{figure_result_position} (c), where the NLOS photon counts from two target bars are isolated from the obscurant and clearly distinguishable despite only separated by about 60 picoseconds. To best assess the capability of the NGSPD in locating the obscured targets, we use two 5 mm wide bars whose width are smaller than the spatial resolution of the system. This width minimizes "long tail" in the histogram attributed to the late arriving photon back-scattered off the target. Also with the narrow bar width, the first returning photon counting peaks can be identified for estimating the nearest distance from the bars to the diffuser with minimal ambiguity \cite{Kirmani58}. In the meanwhile, the simple coaxial single transceiver setup allows us to have same scanning point on the diffuser for both illuminating and photon capturing, providing simpler spherical geometry for the light path between the diffuser and the object rather than ellipsoidal geometry \cite{Gariepy2015}. Given that $\textbf{r}_{\textbf{di}}$ the $i^{th}$ scanning point where $d$ denotes diffuser, $\textbf{r}_{\textbf{oj}}$ as the position of the $j^{th}$ object, the arrival time of the back-scattered photons at each scanning point is simply $t_{i}= \frac{2(\|\textbf{r}_{\textbf{oj}} -\textbf{r}_{\textbf{di}} \|_{2})}{c}$ which denotes the round-trip time-of-flight from the $i^{th}$ scanning point to the bar.

The measured arrival time of the first-photon in the time-resolved histogram at the $i^{th}$ scanning point indicates the round-trip time-of-flight of the NLOS photons between the MEMS mirror and the object via this scanning point. Thus it is first corrected to compensate the optical path difference from transceiver to each scanning points on the diffuser (see supplementary 3). The corrected time-of-flight $t_{ei}$ (e denotes experiment) is then the true round-trip time-of-flight from the $i^{th}$ scanning point to the object, which is later used for retrieving the object position. The top view of the experiment setup is shown in Fig.\ref{figure_result_position} (a), with the position of the object retrieved on the x-z plane. Assuming the object position to be $(x, z)$, we can simulate and map out the arrival time $t_{i}$ of first returning photon from each position $(x,z)$ for every scanning point $\textbf{r}_{\textbf{di}}$. By matching the measured first photon arrival time $t_{ei}$ in the experiment against the $t_{i}$, the position of the bar can be retrieved by a simple least-sum-square evaluation. The sum-square of the error at all the $N$ scanning points is \[ err(x, z) = \sum_{i=1}^{N} \|t_{ei}-t_{i}(x, z)\|_{2}^{2}\] for a given coordinate $(x,z)$ on the plane. In this evaluation, the ensemble of probable position for the object $\textbf{r}_{\textbf{oj}}$ retrieved from one scanning point $\textbf{r}_{\textbf{di}}$ forms a spherical surface centered at $\textbf{r}_{\textbf{di}}$ with radius $\frac{c t_{ei}}{2}$. With $N$ scanning points, $N$ probability distribution spheres are defined. The $(x,z)$ point with least sum-square-distance to all the spheres, or minimum $err(x,z)$, gives the most probable position for the object. Simple geometry of first returning photon is due to coaxial single transceiver setup compared with the separate receiver case \cite{tsai2017geometry}. We use the joint probability density \cite{Gariepy2015} of the least-sum-square to approximate the position of the two bars. Since the NGSPD system has a 10 ps Gaussian-like FWHM of impulse response, the joint probability density is approximated in Gaussian form as \[ P(x,z) = \prod_{i=1}^{N} e^{-\frac{\|t_{ei}-t_{i}(x, z)\|_{2}^{2}}{2\sigma_{t}^{2}}} = e^{-\frac{err(x,z)}{2\sigma_{t}^{2}}}, \] where $\sigma_{t}$ is the standard deviation of the time-resolved measurement and approximated to be FWHM$/2=5$ps. The joint probability of the object position on x-z plane are labeled in Fig.\ref{figure_result_position}(b) which is in 0.5 $\times$ 0.5 mm resolution, where the highest probability reveals the exact locations of the bars. This NLOS positioning retrieval requires only time-of-flight information to reach millimeter resolution, thanks to the advantage of NGSPD in negating undesirable photons. Naturally, the highly resolved photon counting histogram acquired via NGSPD also allows distinguishing the surface normal of the obscured bars, which is shown in supplementary note 4.

\section*{NLOS acousto-optics sensing}

Being able to optically gating the information-carrying NLOS photons off the obscured NLOS target with few picosecond resolution, the NGSPD provides a straightforward method of capturing the quickly diminishing informative photon in NLOS scenario. This capability provides a new NLOS detection modality of NLOS optical sensing in complex environment. Comparing with many existing NLOS detection methods, the NGSPD can directly retrieve the vibration information of the hidden object.

We perform a proof-of-principle acousto-optics sensing on obscured NLOS target via non-interferometric single-photon counting vibrometry \cite{2021arXiv210304771R} based on sampling of photon count while gating the photons from the target. The single-photon counting based vibrometry captures the acoustic signals by continuously sampling the detected photon counts over a fixed dwell time using FPGA (1 kHz sampling rate in this case). Then we apply short-time Fourier Transform to the generated time series of photon counts to obtain the spectrogram which reveals the acoustic signals. The experiment setup is identical to Fig.\ref{figure_result_position} (a), where the 2 bars are excited separately at 2 different vibrating frequency by using two cellphones. The two cellphones playing sound wave at constant but different frequencies are actuating the two bars by simply leaning on each mounting base. The probe beam is pointed on a fix scanning point on the diffuser, where the time-resolved measurement capturing the photon counting peaks originated from two bars at different temporal positions. As the time-of-flight locations of the bars were identified in the previous section, the NGSPD detects back-scattered photons of one single bar by temporally setting the pump at corresponding delay. Thus, acoustic vibration signal from the same bar is captured via time series of photon counting measurement with a preset dwell time, while photons from the other bar is temporally gated out. At this scanning point, the time-of-flight difference of the first-arriving peak of the two bars is 60 ps as observed in Fig.\ref{figure_result_position} (c). 

\begin{figure}[!t]\center
\includegraphics[width=15.0 cm]{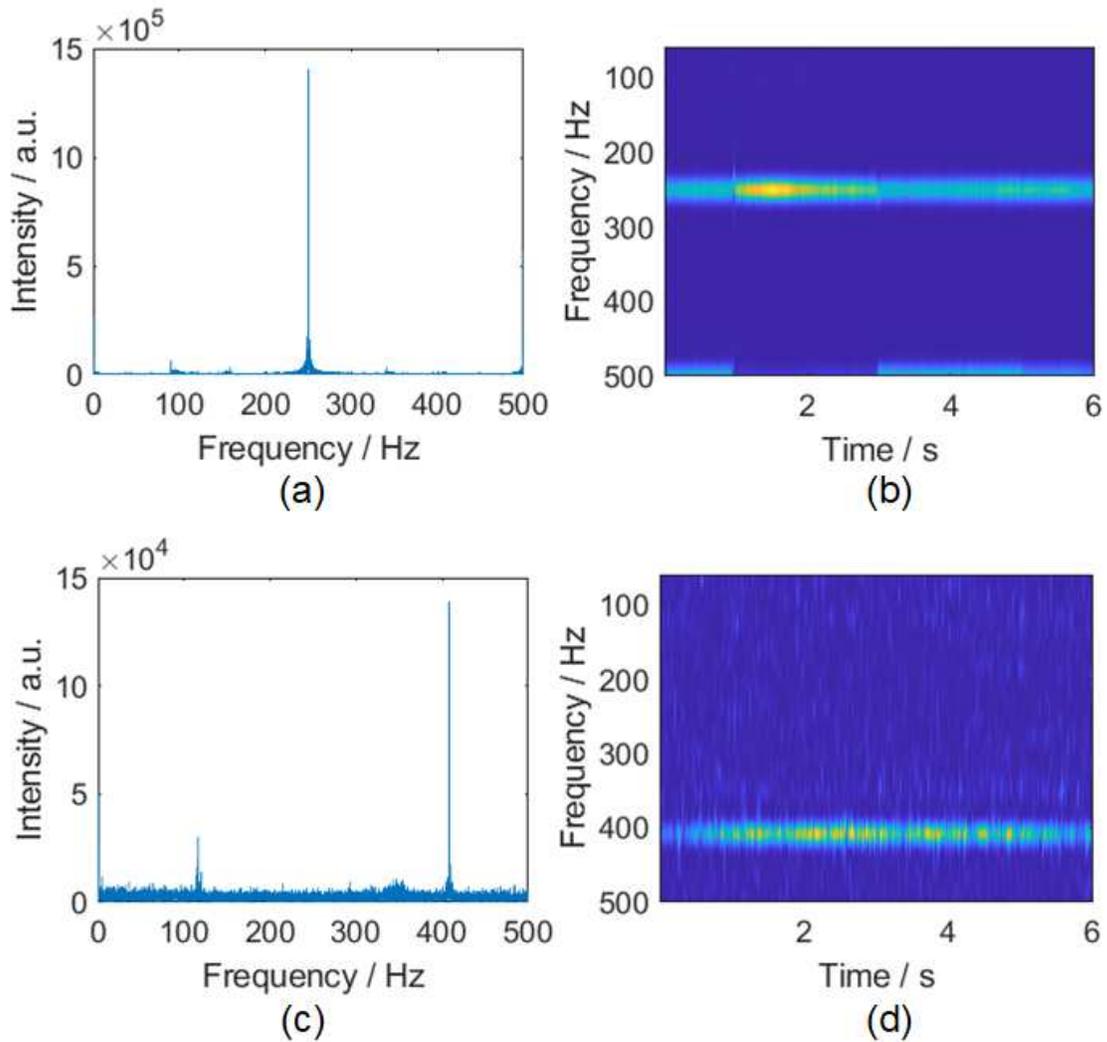}
\caption{\textbf{Results of NLOS acoustic sensing} Acoustic information retrieved from the two bars under the sampling rate of 1 kHz. Figure (a) and (b) shows the Fourier Transform and spectrogram of the 250 Hz acoustic signal from one bar. The second harmonic response of the actuation shows up at the bottom of (b). The unevenness of the frequency response is due to the vibration-caused speckle perturbation of the back-scattered light. Figure (c) and (d) gives the 420 Hz signal actuated on the other bar.}
\label{figure_result_acous}
\end{figure}

The vibration signals from the two bars are isolated, as demonstrated in Fig.~\ref{figure_result_acous}. In each spectrogram, only one actuation frequency is manifested which highlights another advantage of NGSPD on targeted NLOS acoustic-optics sensing with high selectivity and spatial resolution \cite{lindell2019acoustic, Doktofsky2020}. High extinction isolation of undesirable photons is enabled by the picosecond temporal gating and single-mode fiber transceiver that captures very few photons other than those from intended target. Note that, one can observe the frequency noises at 120 Hz due to the power line frequency supplied to the ambient LED lighting, and at 335 Hz due to the resonant frequency of the MEMS mirror in the Fourier Transform figures. 

\section*{Discussion}

Existing techniques for NLOS imaging \cite{liu2019non, liu2020phasor, O'Toole2018} and tracking \cite{Gariepy2015, chan2017non} are overly restrictive for practical uses, and rely heavily on the prowess of data post-processing. By nonlinear optical gating and single photon detection, we have demonstrated a novel approach that achieves picosecond single-photon time gating while rejecting orders of magnitude stronger background noise. It eliminates the otherwise detrimental detection piling-up effects \cite{Heide2018} and allows coaxial NLOS measurement to provide direct time-of-flight information of hidden objects. As such, hidden NLOS scenes, even those additionally occluded, can be reliably reconstructed at centimeter resolution, releasing the need for intense computational imaging or complicated, scene-specific propagation models \cite{Thrampoulidis2019}. The same approach also enables non-interferometric NLOS acousto-optics sensing capable of locating hidden objects by their vibrational frequencies. These results highlight the prospect of hybrid or cross-modality NLOS imaging and sensing, by applying far-reaching acoustics waves to excite objects around the corner and using NLOS single photon detection to read the acoustic response  \cite{lindell2019acoustic}. One major drawback of the current NGSPD approach is the need of temporally delay the gating pump pulse for retrieving photon arrival time information, which makes the data acquisition time-consuming and limits the imaging depth. Several improvements can be applied to decrease the data acquisition time. For example, multi wavelength phase matching enables up-converting two or even more wavelength bands in one waveguide \cite{chou1999multiple}, which will reduce the acquisition time by times of the multi peak numbers. On the other hand, using a synchronized pump pulse train with higher repetition rate and combined with a correlated time tagger \cite{LaManna:20} for acquiring the macro arrival time of triply-bounced photons, the maximum imaging and sensing depth of the NGSPD system can be improved significantly. 

With the above advantages, this NGSPD system can perform NLOS imaging and sensing over realistic, complex environment, including those of obscured and partially occluded objects, yet without complex reconstruction models. Meanwhile, the nonlinear gated single photon detection presents a new optical measurement modality for various potential NLOS applications in imaging, sensing, and communications \cite{2019LSA.....8...69C,Sajeed2021}. An interesting future study of this NLOS imaging technique is to exploit pristine and picosecond-resolved photons arrival time histogram for reconstructing NLOS spatial information with only single illumination point aided by machine learning \cite{Turpin:20}, which is expected to significantly improve its functionality and imaging speed.



\section*{Author Contributions}

S.Z., Y.M.S., P.R., and Y.H. contributed extensively to the work presented in this paper.

\section*{Additional Information}
\subsection{Competing interests:}The authors declare no competing interests.
\section*{Correspondence} Correspondence and requests for materials
should be addressed to (email: ysua@stevens.edu and yhuang5@stevens.edu).

\begin{methods}
\subsection{Picoseconds Non-linear Optical gating} The nonlinear optical gating NLOS system utilizes the sum-frequency of two picosecond pulse trains, where one is the pump(6.6ps FWHM, 1565.5 nm) and the other is the outgoing probe(6.1ps, 1545.1 ps). The two pulse trains are generated by carving a mode lock laser(50MHz) using a pair of cascaded 200 GHz dense-wavelength-division multiplexing (DWDM) filters for reach frequency, thus the two pulse trains are nearly transform limited. The temporal intensity and phase profile of the two pulse trains are measured using a frequency resolved optical gating (FROG) pulse analyzer to quantify the temporal gating width(See Supplementary Note 1). The pump is used as optical gating that only up-converts the signal at certain temporal-frequency (TF) mode effectively, which passed through an programmable optical delay line for temporal scanning. The probe is first amplified by an EDFA to about 0.2 nanoJoule per pulse and then transmits out through the transceiver - a free space fiber coupler - and is collimated into a 2.2 mm Gaussian beam. To reduce the internal reflection, the transceiver consists of an angle-polished single mode fiber coupled out from an aspheric lens. The outgoing probe is shot to a MEMS mirror for steering, and is collimated with the MEMS at rest. From the MEMS, the probe first hits the 2-inch diameter metallic diffuser, then propagates to the object and back-scattered to the diffuser, few of the signal photons finally scatter back reversely and couple into the transceiver again. The diffuser is fixed on a rotational stage to measure the relative angle between the its normal and the probe beam. The distance between the MEMS mirror and the diffuser is 90 cm, and the tilt angle between the normal of the diffuser and the collimated probe beam(MEMS at rest) is 20$^{\circ}$. The triply-bounced information-carrying photons come back to the transceiver and get separated from the probe by a fiber circulator with a minimum isolation ratio of 55 dB. The residue probe from the circulator and the information-carrying photons are temporally separated thanks to the narrow optical gating. The information-carrying photons will then be recombined with pump via another DWDM and subsequently fiber-coupled into the NGSPD. The NGSPD is composed of a commercial periodically poled lithium niobate nonlinear waveguide module and a silicon SPAD ($ {\sim}70\%$ efficiency at 780 nm). The signal photons are up-converted to sum-frequency photons in the waveguide, whose center phase matching wavelength is 1559.8 nm and internal conversion efficiency of the up-conversion waveguide is $121\% /(W \dot{} cm^{2})$. The information-carrying signal photons can be effectively upconverted into sum-frequency photons only if they are 1) temporally aligned with the pump pulse; 2) lying in the phase matched wavelength for the pump; and 3) in the fundamental TF mode of the pump. The residue Raman noise is filtered using a narrow band thin film filter after the waveguide. The sum-frequency photons are then detected by the free-running silicon SPAD. A field-programmable-gate-array (FPGA) controls the steering of the MEMS mirror, the ODL and collecting the photon count signal from the SPAD as the central processor.

\end{methods}

\section*{Data availability}
The data that support the findings of this study are available from the corresponding author upon reasonable request.

\clearpage

\clearpage

\section*{Supplementary}

\begin{figure}
\includegraphics[width=\linewidth]{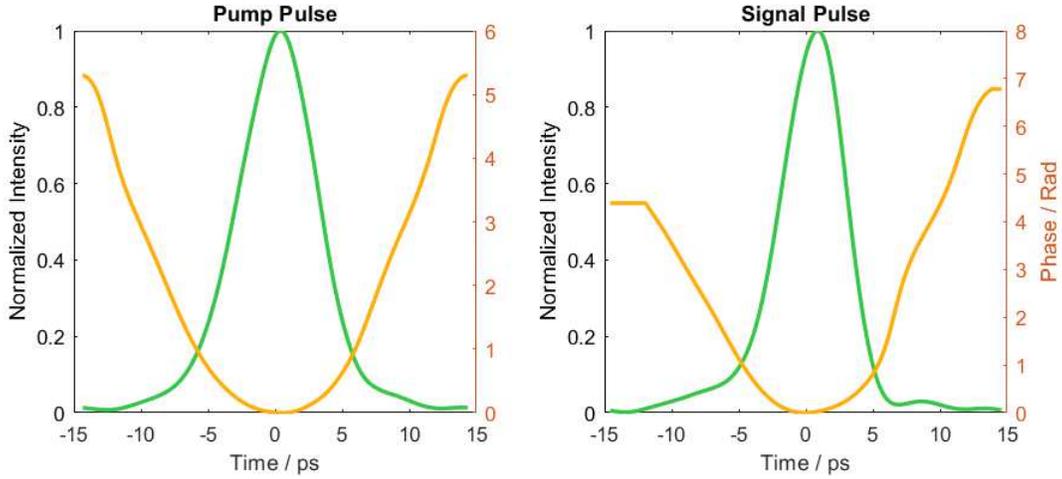}
\caption{Retrieved pulse shapes by the Frequency Resolved Optical Gating (FROG). Left figure and right figure show amplitude and phase profile of generated pump and signal (outgoing probe) pulses carved from the MLL at 1565.5 nm and 1554.1 nm, respectively.} \label{fig_pumpsig}
\end{figure}

\begin{figure}
\centering
\includegraphics[width=0.6\linewidth]{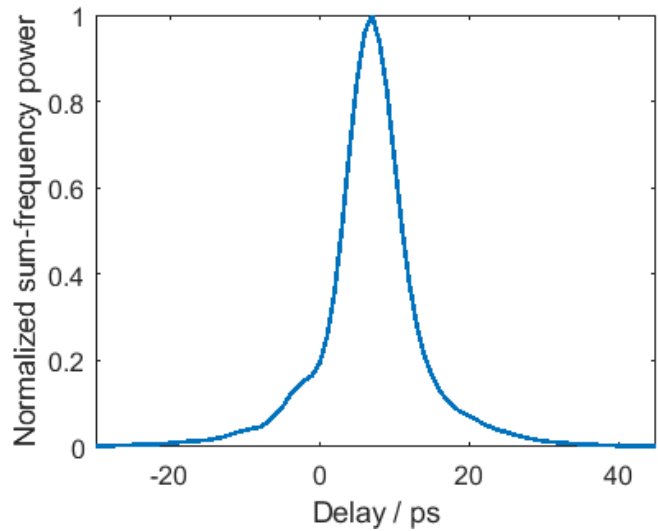}
\caption{Temporal resolved measurement of the impulse response of the sum-frequency signal. The FWHM of the response is 10 ps.} \label{fig_sf}
\end{figure}

\begin{figure}
\centering
\includegraphics[width=0.8\linewidth]{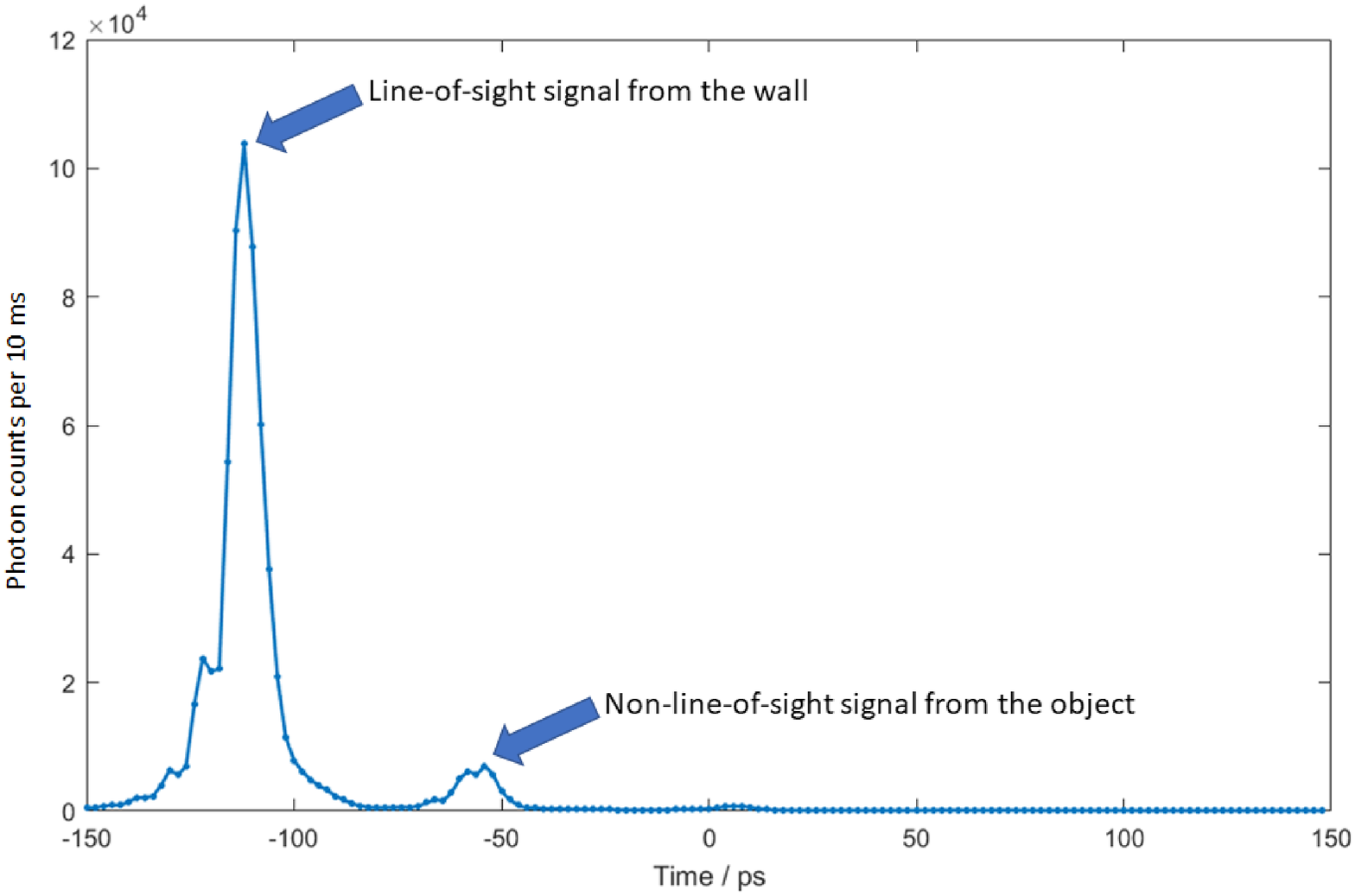}
\caption{Temporal resolved measurement of the visible surface and the hidden surface. The peak at -112 ps is the back-scattered signal from the visible surface, and the bump of the photon counts at around -58 ps comes from the hidden surface, which experienced three bounces. The time here indicates the relative delay of the pump, controlled by the optical delay line.} \label{fig_distinguish}
\end{figure}


\subsection{Supplementary Note 1: Pulse shape}

A mode lock laser is filtered using dense wavelength division multiplexer (DWDM) to generate the transform-limited pump and signal pulse. The carved pump and signal pulse are measured using frequency resolved optical gating (FROG), as shown in Fig. \ref{fig_pumpsig}. The FWHM of pump pulse is 6.6 ps, and the FWHM of signal pulse is 6.1 ps. The back-scattered signal photons are up-converted by the pump in the waveguide and generates sum-frequency photons in a pump-undepleted regime. We measure the sum-frequency temporal resolving profile of the signal pulse by sweeping the optical delay line, and the FWHM of the impulse response is 10 ps, as shown in Fig.\ref{fig_sf}.

The narrow temporal gating of our system can be shown in an extreme case, that the hidden surface is put very near to the wall. Here, two piece of metal block are put face to face, only about 1 cm distance between each other. Both the one-bounced photons from the visible surface and the much weaker three-bounced photons from the hidden surface are received by the transceiver, and the time-histogram shows that the system can distinguish and isolate the visible and the hidden surface, as shown in Fig.\ref{fig_distinguish}.

\subsection{Supplementary Note 2: Noise Analysis and external noise tolerance}

The background noise photon counts of the imaging system are mainly attributed to the noise from the nonlinear gated single photon detector(NGSPD). The NGSPD noise consists of the intrinsic dark count rate of the Si-APD (100 Hz) and the noise of the upconversion module (2000 Hz). Two Raman scattering processes predominantly generate noise photon counts on the sum-frequency band in the upconversion module: (i) pump photons Raman scatterring into the signal band (centered at 1554.1 nm) then upconverting with the strong pump (centered at 1565.5 nm) via SFG, (ii) the Raman scattering of the SH light created by the pump. Operating the system at unity conversion efficiency with pump peak power of about 0.7 W (220 $\mu$W average power), the Raman noise photon count is about 2000~Hz, giving a total dark count rate of 2100Hz per voxel. As we are using an EDFA to amplify the probe power, a tiny portion of the amplified spontaneous emission (ASE) power are reflected by the circulator (-55 dB isolation) and goes into the NGSPD. However, the ASE occurs in the full temporal domain while the NGSPD only effectively detects a narrow temporal window within the whole period( ~10 ps in 20000 ps period), such that the background count caused by the residue ASE is greatly suppressed. Summing the noise source together, this corresponds to low noise probability of 4.2$\times 10^{-5} $ per pulse per delay point due to the single detection mode of NGSPD\cite{Sua2017}.

Besides the constant background noise from the NGSPD, our system rejects external noise photon counts far better than conventional single photon detection. By inserting a noise source (the amplified spontaneous emission(ASE) noise of another EDFA, filtered using the same filter as the probe pulse) which has identical spectral distribution of the signal, our NGSPD shows 36 dB higher noise rejection than a 1-ns gating InGaAs detector\cite{rehain2020noise}. Thus the external noise count can be neglected, and the background noise can be treated as a constant number. The retrieved temporal signal counts $y(t)$ can be treated as $y(t)=\mathcal{P}(Ax(t)+e)$, where $A$ is the impulse response, $e$ is the background noise level, $x(t)$ is the reflection distribution from the hidden object, and $\mathcal{P}$ is Poisson distribution. Currently the signal photon counts at certain temporal delay is about 10 times higher than the background noise, so the photon count fluctuation is much lower than the signal itself. In the pre-process for NLOS imaging, we only consider $y(t) \approx Ax(t)+e$ to retrieve a filtered data, in order to get rid of the background noise and to obtain a finer temporal data for image reconstruction. 

\begin{figure}
\centering
\includegraphics[width=\linewidth]{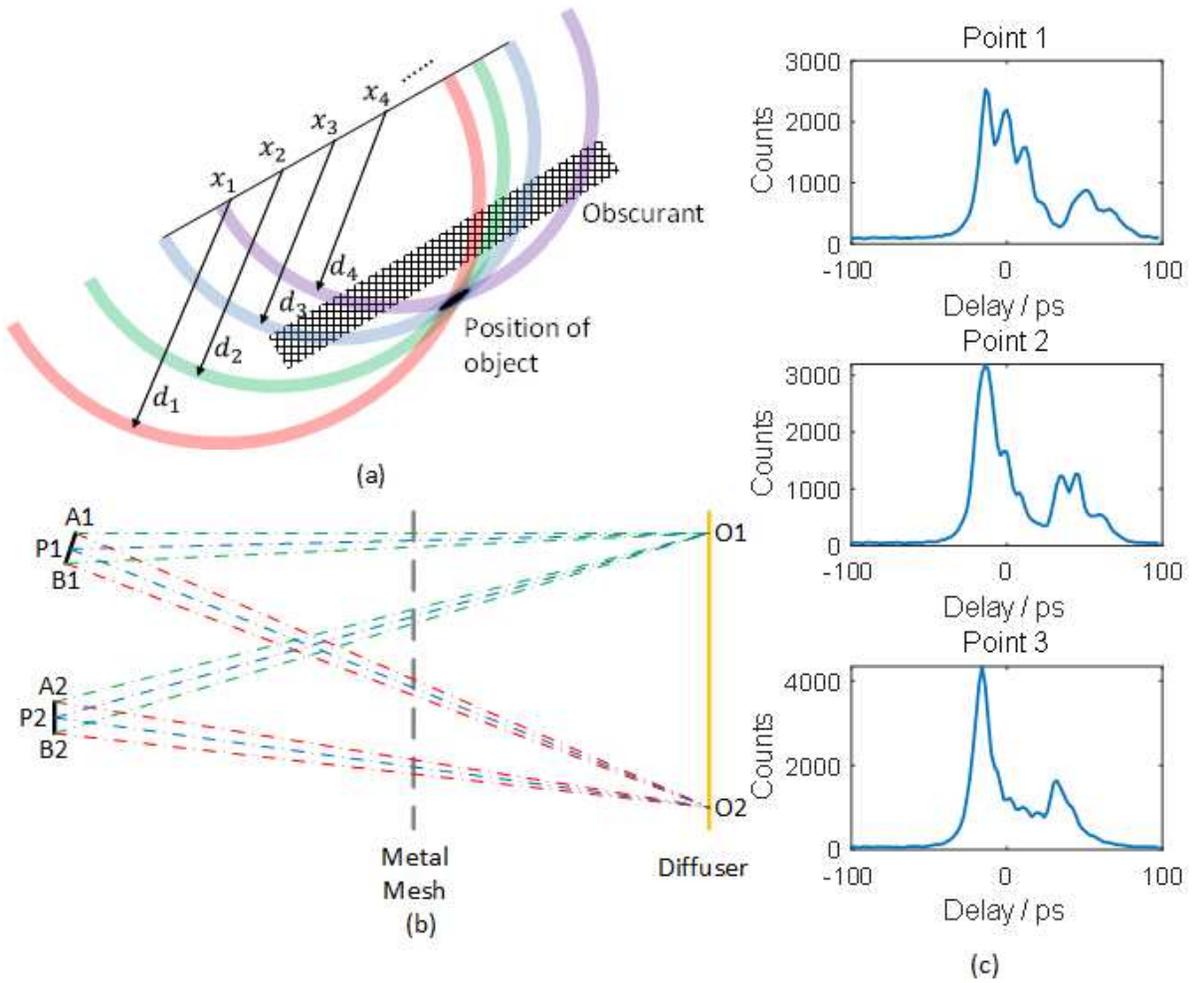}
\caption{NLOS positioning. (a) depicts the spherical probability distribution of the NLOS positioning experiment. The time-resolved measurement at each scanning point $x_{i}$ provides a distance $d_{i}$ to the object. (b) shows the geometric of the positioning, O1 and O2 are the scanning points on the diffuser, A1B1 and A2B2 are the bars. (c) is the scanning raw data of the two bars. When scanning point shifts, the time-of-flight of the back-scattered photons from the two bars varies. We simply picked the first peak of the two responses for processing.} 
\label{fig_tilt}
\end{figure}

\subsection{Supplementary Note 3: Analysis of the errors in NLOS positioning}

In our current NLOS setup, the outgoing probe beam reaches the object via the MEMS mirror and the diffuser. The diffuser has a $20^{\circ}$ angle between its surface normal to the direction of the probe beam when the MEMS mirror is at rest. Thus, the time-of-flight from the MEMS mirror to different scanning points on the diffuser are different and need to be corrected. The distance between the MEMS to the $i^{th}$ scanning point on the diffuser can be expressed as $d_{i} = \frac{d}{\cos{\gamma} \cos(\beta) (1 - \tan{\beta}\tan{\alpha})} $, where $d$ is the distance between the MEMS mirror and the diffuser (probe beam at zero tilt angle when MEMS at rest), $\alpha$ is the tilt angle between the normal of the diffuser and the outgoing probe beam, $\beta$ is the yaw angle and $\gamma$ pitch angle of the MEMS mirror at the $i^{th}$ scanning point. Then the distance differences between the scanning point in the center of the diffuser and $i^{th}$ scanning point are compensated by simply shifting the temporal resolved measurement by the time-of-flight difference $t_{i\_adj} = \frac{d_{i}-d_{center\_pixel}}{2c}$.

The first returning signal peak of each pixel is picked as the first-arrival signal photons from the bar. Although the width of the bars are less than the spatial resolution of the system, the time of flight is still different for the two edges of the bar. Considering the geometry of the setup as Fig. \ref{fig_tilt}b shows, the time of flight uncertainty on the $i^{th}$ scanning point for the $j^{th}$ bar can be interpreted as $ ds = |\overline{O_{i}A_{j}}-\overline{O_{i}B_{j}}|=\sqrt{\overline{O_{i}P_{j}}^{2} + (\frac{\overline{ A_{j}B_{j} }}{2})^{2} - \cos{\angle{O_{i}P_{j}A_{j}}} \cdot \overline{A_{j}B_{j}} \cdot \overline{O_{i}P_{j}} } - \sqrt{\overline{O_{i}P_{j}}^{2} + (\frac{\overline{ A_{j}B_{j} }}{2})^{2} + \cos{\angle{O_{i}P_{j}A_{j}} } \cdot \overline{A_{j}B_{j}} \cdot \overline{O_{i}P_{j}}}$, where $\overline{A_{j}B_{j}}$ is the bar width (about 5mm), and $\overline{O_{i}P_{j}}$ is the distance from the scanning point to the midle point of the bar (about 12 cm). The complementary of the tilt angle, $\angle{O_{i}P_{j}A_{j}}$,  is different at each scanning points. In the experiment, the two bars were facing at the center of the wall, such that the largest tilt angle is about $12^{\circ}$ ($\angle{O_{i}P_{j}A_{j}}$ about $78^{\circ}$), which corresponds to ${ds}_{max} \approx 1mm$ or $6.7ps$ temporal difference. 

We used a simple least square approximation for the bar positioning, which does not assume a specific shape of the surface. So the largest error of one scanning point can be $\Delta t = \sqrt{10^2 + 6.7^2} \approx 12 ps$. By increasing the number of pixels $N$ which is 16 in current experiment, we can decrease the standard deviation of the result by $\sqrt{N}$, which is shown as the very narrow depth distribution in the positioning result.

\begin{figure}
\centering
\includegraphics[width=\linewidth, height=5.3 in]{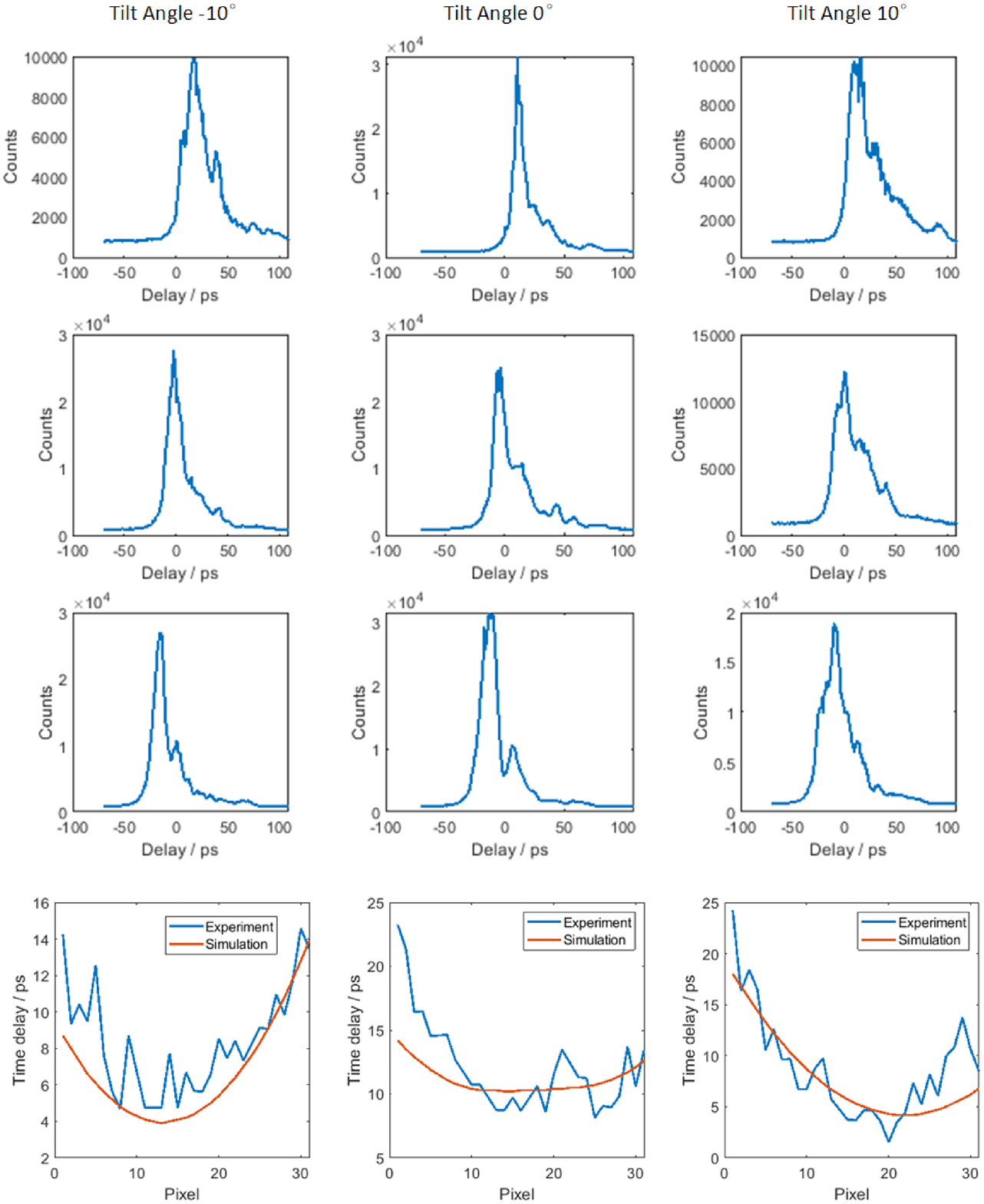}

\caption{NLOS orientation measurement for different tilt angle at -10$^{\circ}$, 0$^{\circ}$ and 10$^{\circ}$. The first three rows of plotting are original data, each row corresponds to one pixel and each column corresponds to one tilt angle. At each pixel, we pick the time of the earliest arriving signal peak, and compare with the simulation result at the same scanning points, which are plotted in the last row. For the pixels on the edge, the error from the experiment and simulation differs most, which can be brought in by the fact that the intensity from the edge is lower than other parts, then the real earliest peak can be immersed.} 
\label{fig_error}
\end{figure}

\subsection{Supplementary Note 4: NLOS orientation retrieval}

The time-of-flight sensitivity of NGSPD provides the potential of evaluating the normal of the hidden surface just using the time-of-flight of the first-arrival photons. A 1.2-cm width bar (should be wider than the spatial resolution) taped with retroreflector is used as the object, and is put onto a rotational stage in front of the diffuser. First, its surface normal is adjusted to be parallel to the surface normal of the diffuser, which is labelled as $0^{\circ}$. Then the bar is yawed using the rotational stage. We measure the time-resolved histogram of the bar on a row of scanning points, at each of the three yaw angles($-10^{\circ}$, $0^{\circ}$ and $10^{\circ}$). The earliest returning peak on the time-resolved histogram is chosen and plotted for each scanning point as shown in the last row in Fig.\ref{fig_error}. The simulation results of the first returning photons' arrival time are plotted together, which shows promising consistency comparing with the experiment results. Thus the travel time of the first returning photons reveals the surface normal of the object. The resolving algorithm for normal angle evaluation needs to be developed in further steps. 

\subsection{Supplementary Reference}

\end{document}